\documentclass[12pt]{article}
\usepackage{}
\usepackage{epsfig}

\usepackage{a4}
\usepackage{amsmath}

\usepackage{epsfig}
\usepackage{amsmath}
\usepackage{amsfonts} 
\usepackage{amssymb}
\usepackage{ifthen}

\begin{document}
\title{On the effects of the Chern-Simons term in an Abelian
\\
 gauged Skyrme model in $d=4+1$ dimensions}
\author{{\large Francisco Navarro-L\'erida,}$^{1}$
{\large Eugen Radu}$^{2}$
and {\large D. H. Tchrakian}$^{3,4}$ 
\\ 
\\
$^{1}${\small Dept.de F\'isica Te\'orica and IPARCOS, Ciencias F\'isicas,}\\
{\small Universidad Complutense de Madrid, E-28040 Madrid, Spain}\\ 
$^{2}${\small Departamento de Fisica da Universidade de Aveiro and }\\
{\small Center for Research and Development in Mathematics and Applications,}
\\
 {\small 
 Campus de Santiago, 3810-183 Aveiro, Portugal
}\\ 
$^{3}${\small School of Theoretical Physics, Dublin Institute for Advanced Studies,}
\\{\small 10 Burlington Road, Dublin 4, Ireland }
\\
  $^{4}${\small  Department of Computer Science, NUI Maynooth, Maynooth, Ireland}}

\date{\today}
\newcommand{\dd}{\mbox{d}}
\newcommand{\tr}{\mbox{tr}}
\newcommand{\la}{\lambda}
\newcommand{\om}{\omega}
\newcommand{\ka}{\kappa}
\newcommand{\ta}{\theta}
\newcommand{\f}{\phi}
\newcommand{\vf}{\varphi}
\newcommand{\vr}{\varrho}
\newcommand{\F}{\Phi}
\newcommand{\al}{\alpha}
\newcommand{\bt}{\beta}
\newcommand{\ga}{\gamma}
\newcommand{\de}{\delta}
\newcommand{\si}{\sigma}
\newcommand{\Si}{\Sigma}
\newcommand{\za}{\zeta}
\newcommand{\bomega}{\mbox{\boldmath $\omega$}}
\newcommand{\bsi}{\mbox{\boldmath $\sigma$}}
\newcommand{\bchi}{\mbox{\boldmath $\chi$}}
\newcommand{\bal}{\mbox{\boldmath $\alpha$}}
\newcommand{\bpsi}{\mbox{\boldmath $\psi$}}
\newcommand{\brho}{\mbox{\boldmath $\varrho$}}
\newcommand{\beps}{\mbox{\boldmath $\varepsilon$}}
\newcommand{\bxi}{\mbox{\boldmath $\xi$}}
\newcommand{\bbeta}{\mbox{\boldmath $\beta$}}
\newcommand{\ee}{\end{equation}}
\newcommand{\eea}{\end{eqnarray}}
\newcommand{\be}{\begin{equation}}
\newcommand{\bea}{\begin{eqnarray}}

\newcommand{\ii}{\mbox{i}}
\newcommand{\e}{\mbox{e}}
\newcommand{\pa}{\partial}
\newcommand{\Om}{\Omega}
\newcommand{\vep}{\varepsilon}
\newcommand{\bfph}{{\bf \phi}}
\def\theequation{\arabic{equation}}
\renewcommand{\thefootnote}{\fnsymbol{footnote}}
\newcommand{\re}[1]{(\ref{#1})}

\newcommand{\eins}{1\hspace{-0.56ex}{\rm I}}
\newcommand{\R}{\mathbb R}
\newcommand{\C}{\mathbb C}
\newcommand{\p}{\mathbb P}
\renewcommand{\thefootnote}{\arabic{footnote}}

\maketitle


\medskip

\begin{abstract}
 We study an Abelian gauged $O(5)$ Skyrme model in $4+1$ dimensions, 
featuring the $F^4$ Maxwell and the Chern-Simons terms.
Our aim is to expose the mechanism, discovered in the analogous Abelian gauged $O(3)$ Skyrme model in $2+1$ dimensions, 
which leads to the {\it unusual}
relation of the mass-energy $E$ to the electric charge $Q_e$ and angular momentum $J$, and, to the change in the value of the 
``baryon number'' $q$ due
to the influence of the Abelian field on the Skyrmion. Chern-Simons dynamics together with the dynamics of the gauged Skyrme scalar, allows for solutions
with varying asymptotic values of the {\it magnetic} field, resulting in these {\it unusual} properties listed. Numerical work is carried out on an effective one
dimensional subsystem resulting from imposition of an enhaced radial symmetry on $\R^4$.

\end{abstract}
\medskip

\section{Introduction}

The Skyrme model has started  almost sixty years ago
\cite{Skyrme:1961vq,Skyrme:1962vh},
providing the very first
explicit example of solitons in a relativistic non-linear field theory in $d=3+1$
spacetime dimensions.
This model has been generalised subsequently to all~\footnote{These are a direct extensions of
the usual Skyrme model~\cite{Skyrme:1961vq,Skyrme:1962vh}, namely
of the $O(4)$ sigma model on $\R^3$. $O(D+1)$ Skyrme models on $\R^D$ consists of all possible
kinetic terms and potential term, consistent ,with the Derrick scaling requirement. Such models are analysed in Apendix {\bf B} of
Ref.~\cite{Tchrakian:2015pka}, and in references therein.
} spacetime dimensions. Concrete Skyrmions were constructed in \cite{Piette:1994jt}
for the $O(3)$ sigma model on $\R^2$, and for the $O(5)$ model on $\R^4$ in
\cite{Brihaye:2017wqa}. These are finite energy solutions  
(Skyrmions) whose energy is bounded from below by a topologically invariant charge $n$, which is the baryon number.

The situation changes in the presence of gauge fields. 
The gauging prescription given in \cite{Tchrakian:2015pka,Tchrakian:1997sj} provides for a lower bound $q$ for the energy of
the gauged system, $q$ being the volume integral of a density $\vr$ which, as required, is {\it Lorentz} and {\it gauge invariant}.
Like the baryon number density prior to gauging, say $\vr_0$, its gauge-deformed counterpart $\vr$ is by construction
{\it essentially total divergence}, meaning that in a constraint-compliant parametrisation
of the Skyrme scalar it is {\it total divergrence}.
 This property enables the evaluation of the lower bound of the energy as a surface integral determined only
by the asymptotic values of the solution, provided it is regular at the origin. 
The energy lower bound $q$ 
coincides with the topologically invariant baryon number $n$ in the
gauge decoupling limit~\footnote{
This applies to any gauged system that in the gauge decoupling limit supports topologically stable solutions, notably the Goldtone
model~\cite{Tchrakian:2010ar} on $\R^D$ described by the real $D$-component scalar, whose solitons are stabilised by the topological
$winding\ number\ n$. 
After gauging  with $SO(N)$, $2\le N\le D$ using the prescription given for the Skyrme
systems~\cite{Tchrakian:1997sj,Tchrakian:2015pka}, the energy lower bound cannot be saturated~ \cite{Tchrakian:2002ti} and is topologically
invariant only for gauge group $SO(D)$.
},
and can be seen as a deformation of the baryon number $n$. 
However, unlike the baryon number $n$, $q$ is not topologically invariant.
It may nonetheless be reasonable to call $q$ a "gauge-deformed baryon number".

Since this gauging prescription does not preserve the topological invariance of the energy lower bound, the gauge group in question is not contrained
by the topology and hence one can gauge the $O(D+1)$ Skyrme scalar on $\R^D$ with $SO(N)$, $2\le N\le D$ gauge group~\footnote{
In the particular case of $SO(2)$ gauging of the $O(4)$ model on $\R^3$, this prescription coincides with that employed in
Ref.~\cite{Callan:1983nx} much earlier.}.

It is because $q$ is not topologically invariant that there exist finite energy solutions for varying values of $q$, and this is what gives
rise to the features of gauged Skyrmions discussed below.
 
\medskip

The new features in question were first discovered in
the recent work \cite{Navarro-Lerida:2018giv,Navarro-Lerida:2018siw}, for
the Abelian gauged $O(3)$ Skyrme model supplemented with a Chern-Simons term in $d=2+1$ dimensions.
The main new features reported there are:
\begin{itemize}
\item
the change, or deformation, of the baryon number $q$ of the Skyrmion due to the influence of the gauge field through
Chern-Simons dynamics, with
\begin{eqnarray}
\label{q2+1}
q=n+a_\infty ,
\end{eqnarray}
where $a_\infty$ is the asymptotic value of the magnetic gauge potential and $n=1,2\dots$ is the winding number of the Skyrme scalars. 
\item
the $unusual$ dependence of the mass-energy $E$ on the (global) electric charge $Q_e$ and angular momentum $J$ of the Skyrmion,
with $E$ taking both negative and positive
gradients, in contrast with the $usual$ positive slope of monotonically increasing $E\ vs.\ (Q_e,J)$.
The values of $Q_e$ and $J$ are also determined by the magnetic flux at infinity, with
\begin{eqnarray}
\label{QJ2+1}
Q_e=8\pi (n-a_\infty),~~J=4\pi (a_\infty^2-n^2).
\end{eqnarray}
\end{itemize}


The properties  \re{q2+1} and \re{QJ2+1} of $SO(2)$ gauged Skyrmions depend on the existence of solutions with $a_\infty\neq 0$, which occurs exclusively by virtue of
Chern-Simons (CS) dynamics. This limits such phenomena to odd dimensional spacetimes, where CS densities can be defined. Thus for example these
phenomena are absent in an $SO(2)$ gauged Skyrme model in $3+1$ dimensions,
where there is no CS term,  with the mass-energy of the solutions increasing monotonically with increasing
electric charge, and, the topological charge $q$ remaining fixed \cite{Navarro-Lerida:2018giv}, in which case $a_\infty=0$
and $J\sim Q_e$.

\medskip

The purpose of the present work is to demonstrate that the mechanism giving rise to the effects \re{q2+1} and \re{QJ2+1}, of the Chern-Simons dynamics in
the Abelian gauged $O(3)$ Skyrme model in $2+1$ dimensions,
are also present in the Abelian gauged $O(5)$ Skyrme model in $4+1$ dimensions. 
Moreover, this mechanism is present in all Abelian gauged $O(D)$ Skyrme models in $D+1$ specetime dimensions (with $D=2p$).

The Skyrme model we employ here is the Abelian gauged $O(5)$ model in $4+1$ dimensions, which is a variant
of the model studied in Ref.~\cite{Navarro-Lerida:2020jft}, 
with the (usual) Maxwell $F^2$ term in the action
replaced here with the $F^4$ ``Maxwell-like'' term. 
This change allows for a slower asymptotic decay of the gauge field, resulting in
a nonvanishing magnetic field at infinity. This is
what enables the construction of solutions exhibiting the promised effects of Chern-Simons dynamics
in $4+1$ dimensions, replicating qualitatively those found in $2+1$ dimensions.
Also, following Ref. \cite{Navarro-Lerida:2020jft}, we shall 
impose an enhanced symmetry on the system that renders the residual system one-dimensional,
depending only on the radial variable, bypassing the more general bi-azimuthal symmetry \footnote{This short-cut is justified since we have verified in \cite{Navarro-Lerida:2020jft} that
qualitative features of the more general solutions  do not differ from those of the enhanced symmetry 
solutions.}.

\medskip

The paper is structured as follows.
In Section {\bf 2}, we present the model and impose symmetry, while in Section {\bf 3}, we present our results. 
In Section {\bf 4}, we summarize our results and make some remarks,
including to point out that the mechanism described here can be extended to all $d=2p+1$ dimensions.

\section{The model and symmetry imposition}

\subsection{The matter content}
The only difference between the $SO(2)$
 gauged Skyrme model in this work  
 and the model studied in \cite{Navarro-Lerida:2020jft} 
is that the quadratic ($p=1$) Maxwell term $F^2$ is replaced
\footnote{Models with
such  higher-order terms have been extensively studied 
in the literature 
(although mainly for the non-Abelian case)
leading to a variety of new features (see Ref. \cite{Tchrakian:2010ar} for a review).
Moreover,  such $F(2p)^2$ terms 
occur in  Born-Infeld theory \cite{Tseytlin:1997csa}
 or in the higher loop corrections to the $d = 10$
heterotic string low energy effective action \cite{Polchinski:1998rr}. 
Here, however, we adopt a 'phenomenological'
viewpoint and replace  $F^2$ with $F^4$ 
primarily for the purpose of reproducing the results in \cite{Navarro-Lerida:2018giv,Navarro-Lerida:2018siw}.
}
 by the
quartic ($p=2$) ``Maxwell-like'' term $F^4$.

Therefore, for a $d=5$ flat spacetime geometry $ds^2=\eta_{\mu \nu} dx^\mu dx^\nu$
(with $x^0=t$-the time coordinate),
 the model's Lagrangian reads
\be
\label{L1}
{\cal L}= \frac{1}{4!}\la_M F_{\mu\nu\rho\si}^2+\ka\,\Om_{\rm CS}^{(5)}-\la_1|\f_{\mu}^{a}|^2+\frac32\la_2|\f_{\mu\nu}^{ab}|^2
-\frac{1}{(3!)^2}\la_3|\f_{\mu\nu\la}^{abc}|^2+\la_0 V[\f^5] \,,
\ee
where 
$F_{\mu\nu}=\pa_{[\mu}A_{\nu]}$ is the Maxwell curvature
and
$F_{\mu\nu\rho\si}= F_{[\mu \nu}F_{\rho\si]}$
is a four form (totally antisymmetrised product of two two-forms $F(2)$).
Also 
 $\Om_{\rm CS}^{(5)}$ is the Chern-Simons (CS) density
\be
\label{CS5}
\Om_{\rm CS}^{(5)}=-\frac12\vep^{\mu\nu\rho\si\la}A_{\la}\,F_{\mu\nu}F_{\rho\si}\,,
\ee
and  $\ka$ is the CS coupling constant. $\lambda_i$ ($i=0,\dots, 3$)
and  $\lambda_M$
are the rest of coupling constants that parameterise the model.
To simplify a number of relations in what follows, we set $\lambda_M=1$
without any loss of generality.
 
It is useful the record that the Chern-Simons term \re{CS5}, which mixes electric and magnetic fields,
reduces, apart from an irrelevant total divergence term, to
\be
\label{CS5stat}
\Om_{\rm static}^{(5)}=-\vep^{ijkl}A_{0}\,F_{ij}F_{kl}\ ,\quad i,j,..=1,2,3,4\,,
\ee
in the static limit.

It is further relevant to point out that while the term $F_{ijkl}^2$ in \re{L1} also consists of both electric and magnetic fields, nonetheless
differs qualitatively from the Chern-Simons density in that its dependence on $A_0$ is through $F_{ijk0}^2$, namely on the second power of $F_{i0}$, while \re{CS5stat} is
linearly dependent on $A_0$.

The notation used in \re{L1} is formally $\f_\mu^a=D_\mu\f^a$, the covariant derivative of the Skyrme scalar $\f^a\ ; \ a=1,2,3,4,5$,
which satisfies the
constraint $|\f^a|^2=1$. Then $\f_{\mu\nu}^{ab}$ and $\f_{\mu\nu\la}^{abc}$ are defined as the totally antisymmetrised products
of $\f_\mu^a$, in terms of which
the Skyrme kinetic terms~\footnote{We have omitted the octic kinetic term $|\f_{\mu\nu\rho\si}^{abcd}|^2$ in \re{L1} as we
limit our considerations up to the sextic one.
} in the Lagrangian \re{L1} are expressed.
Also, $V= 1-\f^5$
is the usual Skyrme potential.

Labelling the $O(5)$ Skyrme scalar as 
$\f^a=(\f^{\al},\f^{A},\f^5)$, with $\al=1,2\ ;\ A=3,4$, the covariant derivatives in \re{L1} are defined by 
\bea
\nonumber
\f_\mu^{\al}&=&D_\mu\f^{\al}=\pa_\mu\f^{\al}+A_\mu(\vep\f)^{\al} ~,
\label{coval'}
\\
\f_\mu^{A}&=& D_\mu\f^{A}=\pa_\mu\f^{A}+A_\mu(\vep\f)^{A} ~,
\label{covA'}
\\ 
\nonumber
\f_\mu^{5}&=&D_\mu\f^5=\pa_\mu\f^5\label{cov3}\, ,
\eea
subject to the constraint
\[
|\f^\al|^2+|\f^A|^2+(\f^5)^2=1\,.
\]
Here $\vep$ denotes the Levi-Civita symbol in each of the two-dimensional subsets of internal indices, $(1,2)$ and $(3,4)$, respectively. More specifically, $(\vep\f)^1 = \f^2 , (\vep\f)^2 = -\f^1 $ and similar for indices $(3,4)$.
 
The Maxwell equations resulting from the variations $w.r.t.$ $A_\la$ are,
\be
\label{Max}
F_{\rho\si}\pa_\mu F^{\mu\la\rho\si}-\frac32\ka\vep^{\la\mu\nu\rho\si}F_{\mu\nu}F_{\rho\si}=j^\la[A_\mu,\f^a]\, ,
\ee
where $j^\la[A_\mu,\f^a]$ is the Skyrme current resulting from the variations w.r.t. $A_\la$ of the Skyrme kinetic terms in \re{L1}.
Restricting, for example, to the quadratic term with $\la_1$ in \re{L1}, one finds
\be
\label{jla}
j_\la=-2\la_1\left[(\vep\f)^\al\,D_\la\f^\al+(\vep\f)^A\,D_\la\f^A\right] \, .
\ee

The Gauss Law equation, namely the time component of \re{Max}, is
\be
\label{Gauss}
F_{jk}\pa_i F^{i0jk}-\frac32\ka\vep^{0ijkl}F_{ij}F_{kl}=j^0[A_\mu,\f^a]\,,
\ee
from which follows the definition 
\footnote{
It should be stressed that this definition of the electric (global) charge is the Paul-Khare~\cite{Paul:1986ix,Hong:1990yh}
definition, which differs from the usual one where the electric charge is read directly
from the asymptotic decay of the electric potential.
Refs.~\cite{Paul:1986ix,Hong:1990yh} pertain to the Abelian gauged Higgs model in $2+1$ dimensions. The corresponding formulation for the
$O(3)$ sigma model is given in Refs.~\cite{Ghosh:1995ze}.
}
of the electric charge as the volume integral in $\R^4$
\be
\label{electch}
Q_e\stackrel{\rm def.}=-\frac{ 1}{8\pi^2}\int j_0\,d^4x=-\frac{ 1}{8\pi^2}\int\left(F_{jk}\pa_i {{F^i}_0}^{jk}
-\frac32\ka\vep^{ijkl}\,F_{ij}F_{kl}\right)d^4x\,.
\ee

In this work we are interested in time-independent, rotating configurations,
in which case
the energy density is the $T_{00}$ component of the stress tensor $T_{\mu\nu}$ pertaining to the Lagrangian \re{L1}, while the angular momentum densities
here are those in the two planes $x_\al=(x_1,x_2)$ and $x_A=(x_3,x_4)$ in $\R^4$.
These are defined in terms of the components $T_{0\al}$ and
$T_{0A}$ components of $T_{\mu\nu}$ respectively.

The definition of the gauge-deformed ``baryon number'' density~\cite{Tchrakian:1997sj,Tchrakian:2015pka} is given by the generic expression
\bea
\vr&=&\vr_0+\vr_1\ ,\quad\vr_1\stackrel{\rm def.}=\pa_i\Om_i[A,\f]\,,\label{top14}
\eea
with 
$\varrho_0 = \vep_{ijkl} \vep^{abcde}\pa_i\f^a\pa_j\f^b\pa_k\f^c\pa_l\f^d\f^e$,
being the (topological) baryon number density, 
which is deformed, or modified, by the addition of the total divergence term $\vr_1=\pa_i\Om_i$,
yielding the total ``baryon number''
\be
\label{vr10}
q= -\frac{ 1}{64\pi^2}
\int\vr\,d^4x.
\ee

For the system at hand, $\Om_i$ is derived in detail in Ref.~\cite{Navarro-Lerida:2020jft} so we just quote it here
\bea
\Om_i&=&2\cdot 3!\,\vep_{ijkl}\,A_l\,\left\{\frac13(\f^5)^3F_{jk}+\f^5\left(\vep^{\al\bt}\pa_j\f^{\al}\pa_k\f^{\bt}
+\vep^{AB}\pa_j\f^{A}\pa_k\f^{B}\right)\right\}\,.\label{Om2A}
\eea

Also, we point out that the definition \re{Om2A} is more general than the example given in \cite{Tchrakian:2015pka}, where only one pair of the five
components $O(5)$ Skyrme scalar on $\R^4$ were gauged with $SO(2)$. In \cite{Navarro-Lerida:2020jft} and here by contrast, two pairs of the Skyrme scalar are gauged. While in
\cite{Navarro-Lerida:2020jft} this was done for a technical reason, namely to enable the imposition of (enhanced) radial symmetry on the bi-azimuthal system, here, the explicit expression
\re{Om2A} is essential for achieving a gauge-deformed ``baryon number'' that differs from the (topological) baryon number.

While the gauge-deformed ``baryon number'' density given in \cite{Tchrakian:2015pka} features only the electric component $A_0$ of the Abelian gauge connection, the more general
prescription resulting from \re{Om2A} can be seen to feature both the electric component $A_0$ and the magnetic component $A_i$ of the Abelian gauge connection. The presence of $A_i$
is crucial for achieving our aim, to change the baryon number.

\subsection{Imposition of symmetry and quantities of interest}
%
We consider the following parametrization of the spatial $\mathbb{R}^4$ part of the background metric
\begin{eqnarray}
\label{Mink}
ds^2=dr^2+r^2( d\theta^2+\sin^2 \theta d\varphi_1^2+\cos^2 \theta d\varphi_2^2),
\end{eqnarray}
where $0\leq r<\infty$ is the radial coordinate,
and $\theta,\varphi_{1,2} $ are coordinates on $S^3$,
with 
$0\leq \theta\leq \pi/2$
and
$0\leq \varphi_{1,2}<2\pi$.

The dependence on the angular variables $\varphi_{1,2}$ can be factorized
for stationary, axially symmetric configurations, 
the generic configuration possessing a dependence on both $r$ and $\theta$. 
However, the recent work \cite{Navarro-Lerida:2020jft} 
has proposed a special Maxwell-Skyrme (axially symmetric) Ansatz 
which results in 
an enhanced symmetry on the system that renders the residual system one-dimensional,
depending only on the radial variable.
Moreover, this Ansatz provides the natural generalization of
that employed in Ref.
\cite{Navarro-Lerida:2018giv,Navarro-Lerida:2018siw}
in the $d=2+1$ case,  
with
\begin{eqnarray}
\label{n11}
&&
\phi^1=\sin f(r) \sin \theta \cos \varphi_1,~~
\phi^2=\sin f(r) \sin \theta \sin \varphi_1,~~
\\
\nonumber
&&
\phi^3=\sin f(r) \cos \theta \cos \varphi_2,~~
\phi^4=\sin f(r) \cos \theta \sin \varphi_2,~~
\phi^5=\cos f(r)  ,
\end{eqnarray} 
for the Skyrme scalars, and
\begin{eqnarray}
\label{n12}
A=A_\mu dx^\mu=a(r)\sin^2 \theta d\varphi_1+ a(r)\cos^2 \theta d\varphi_2 +b(r)dt~,
\end{eqnarray} 
for the Maxwell field (with a magnetic potential $a(r)$
and an electric one $b(r)$),
the angular dependence being factorized.

This restrictive Ansatz greatly reduces the complexity of the system and simplifies the numerical construction
of solutions.
For example, one finds the following effective Lagrangian of the system
\bea
\label{Leff}
r^{-3}L_{\rm eff}&=&
\frac{4 a^2 }{r^4} \left(\frac{a'^2}{r^2}-b'^2\right)+\frac{8\ka }{r^3} (b a'-a b')a
\nonumber
\\
&&+\lambda_1\left[f'^2+\frac{\sin^2f}{r^2}\left[2+(1-a)^2 -r^2b^2\right]\right]
\nonumber
\\
&&+6 \lambda_2\frac{\sin^2 f}{r^2}\left[f'^2 \left[2+(1-a)^2-r^2 b^2 \right] +\frac{\sin^2 f}{r^2} \left[1+2(1-a)^2 -2r^2b^2\right]\right]\nonumber
\\
&&+ \lambda_3   \frac{\sin^4f}{r^4}\left[f'^2  \left[1+2(1-a)^2-2 r^2b^2 \right]+\frac{\sin^2 f}{r^2}[ (1-a)^2-r^2b^2 ]\right]
\nonumber
\\
&&
+\lambda_0 (1-\cos f),
\eea
the contribution of various terms being transparent. 

Next, we evaluate the one dimensional integral resulting from the imposition of (enhanced) radial symmetry on the integral \re{electch} defining the global charge $Q_e$,
\be
\label{electch1}
Q_e=\int_0^\infty\, dr~\frac{d}{dr}\left(2\frac{a^2b'}{r}+3\kappa\,a^2\right)\, .
\ee
 
The total mass-energy, $E$, and angular momenta  $J_{1,2}$
in the two sub-planes of $\R^4$, 
 of a solution are defined as 
  the volume integral of $T_t^t$ and $T_{\varphi_{1,2}}^t$, respectively,
	with
\begin{equation}
\label{MJ}
E=\int d^4 x \sqrt{g}\,T_t^t,~~
J_1=  \int d^4 x \sqrt{g}\,T_{\varphi_{1}}^t,~~
J_2=  \int d^4 x \sqrt{g}\,T_{\varphi_{2}}^t~,
\end{equation}
where $\sqrt{g}=r^3 \sin \theta \cos \theta$ and $J_1=J_2=J$ for the considered Ansatz.  
The $T_t^t$ component of the stress tensor can be read off \re{Leff} by changing the appropriate signs and setting $\ka=0$, so
it is not necessary to display it again,
while the explicit form of the angular momenta densities is
%
\begin{eqnarray}
\nonumber
&
\frac{T_{\varphi_1}^t}{\sin^2\theta} =  
\frac{T_{\varphi_2}^t}{\cos^2\theta} = 
 \frac{8}{r^4}a^2 a'b'
-2\sin^2f (1-a)b
\left[
\lambda_1+6\lambda_2 \left(f'^2+\frac{2\sin^2f}{r^2} \right)
+2\lambda_3  \frac{\sin^2f}{r^2}
\left(
f'^2+\frac{\sin^2f}{2r^2}
\right
)
\right]~.
 \end{eqnarray}  
Using the field equations, one can show that  
  \begin{eqnarray}
	\label{sup1}
 \sqrt{g}T_{\varphi_1}^t=\sin^3\theta \cos \theta~ {\cal S}',~~  
 \sqrt{g}T_{\varphi_2}^t=\cos^3\theta \sin \theta~ {\cal S}',~~
  \end{eqnarray}
	where
  \begin{eqnarray}
	&&
 {\cal S}=
-\frac{8 }{r} a^2(1-a)b'-4\kappa a^2(3-2a),
  \end{eqnarray}
which  makes manifest  the
total derivative structure of $T_{\varphi_i}^t$.

Finally, the one dimensional baryon number density $\vr_0$, and $\vr_1=\pa_i\Om_i$ given by the choice \re{Om2A},
reduce to
\bea
\label{ro5}
\vr_0&=&-\frac{4!}{r^3}\frac{d}{dr}\left[\cos f-\frac13\cos^3f\right]\ \ {\rm and} \ \ \
\vr_1=\frac{4!}{r^3}\frac{d}{dr}\left[\frac23\,a^2\cos^3f+a\sin^2f\cos f\right]\, .
\eea
yielding $\vr=\vr_0+\vr_1$ as $per$ \re{top14}, from which $q$ defined by \re{vr10} can be calculated.


\section{Results}
\subsection{The  boundary conditions and global charges}
The resulting system of three ODEs for the functions
$a,b$ and $f$ 
is solved numerically \footnote{The numerical treatment of the problem was similar to 
that employed in Ref. \cite{Navarro-Lerida:2020jft}.}
subject to the following set of boundary conditions
which are compatible with the finite global charges and regularity requirements:
\begin{eqnarray}
 f(0)=\pi,~~a(0)=0,~~b(0)=b_0,~ f(\infty)=0,~~a(\infty)=a_\infty,~~b(\infty)=b_\infty\, .
\end{eqnarray}
These boundary conditions
are similar to those  in  
\cite{Navarro-Lerida:2020jft},
except for $a(\infty) \neq 0$.
That is,
 for a model with an $F^2$ term in the action,
the contribution of the Maxwell term 
to the total mass-energy diverges unless the magnetic potential
vanishes asymptotically.
This feature is not present when replacing $F^2$ with a $F^4$ (quartic) Maxwell term,
whose faster asymptotic decay allows $a(\infty)$
 to take an arbitrary value \footnote{Note that,
following \cite{Blazquez-Salcedo:2017cqm},
the parameter $a_\infty$ can
be identified with the magnetic 
flux at infinity through the base space $S^2$ of the $S^1$ fibration of  $S^3$,
with the sphere written in term of Euler angles.
Also, it is interesting to remark that several $F^2$-models 
with $a(\infty) \neq 0$ were considered in the literature 
 \cite{Blazquez-Salcedo:2017cqm},
\cite{Blazquez-Salcedo:2017kig}, 
they requiring anti-de Sitter
asymptotics of the spacetime geometry.}.

When replacing these asymptotic behaviour in 
(\ref{electch1}),
(\ref{MJ})
and
(\ref{ro5}),
one finds the simple relations
\begin{eqnarray}
Q_e=3\kappa a_\infty^2,~~
 J=
 4\pi^2\kappa  a_\infty^2(2a_\infty-3),~~q=  1-\frac12 a_\infty^2 ,
\end{eqnarray}
which, as seen from (\ref{q2+1}), (\ref{QJ2+1}),
 are qualitatively similar with those found in the 2+1 dimensional model.

\subsection{Numerical results}
%
We have constructed numerical solutions for several choices of the parameters in the theory,
the profile of a typical solution with 
$a_\infty=0.605$, 
$b_\infty=0.3$
being shown in Fig. \ref{profile}. 

\begin{figure}[h!]
\centering
\includegraphics[height=2.5in]{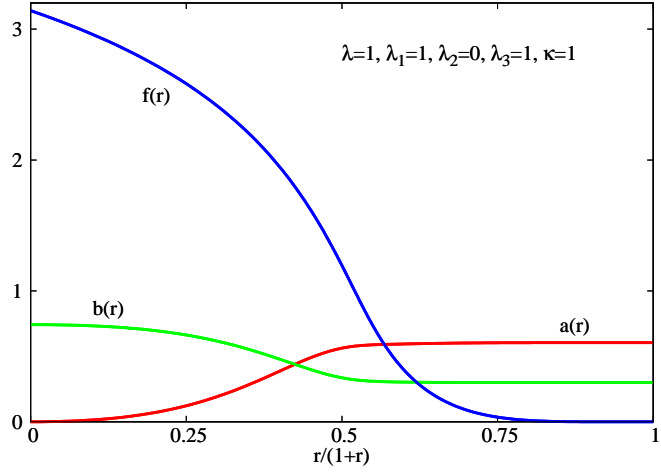}  
\caption{
The profile of a typical SO(2) gauged Skyrme solution is shown as a function of the (compactified)
radial coordinate. 
} 
\label{profile}
\end{figure} 

In what follows, we shall concentrate on the case
 $\lambda=\lambda_1=\lambda_3=1$, 
$\lambda_2=0$, and several values of $\kappa$.
 In Fig.~\ref{fig1} (left panel)
 we represent the asymptotic value 
$b_\infty$ of the electric potential
 $b(r)$ as a function of the asymptotic value $a_\infty$ of the magnetic function $a(r)$.
In contrast to the case in Ref. \cite{Navarro-Lerida:2018giv}, 
$a_\infty$ can only take positive values.
 On the other hand, $b_\infty$ is bounded by the inequality \footnote{This is true for the branches considered in this work. 
However, there are some numerical indications that other branches might exist for which this bound is exceeded.} 
\be
\label{ineq_binf}
b_\infty^2 < \frac{\lambda}{2 \lambda_1} \,.
\ee
The maximum value of $b_\infty$ coincides with the maximum value of $a_\infty$.

\begin{figure}[h!]
\centering
\includegraphics[height=2.8in]{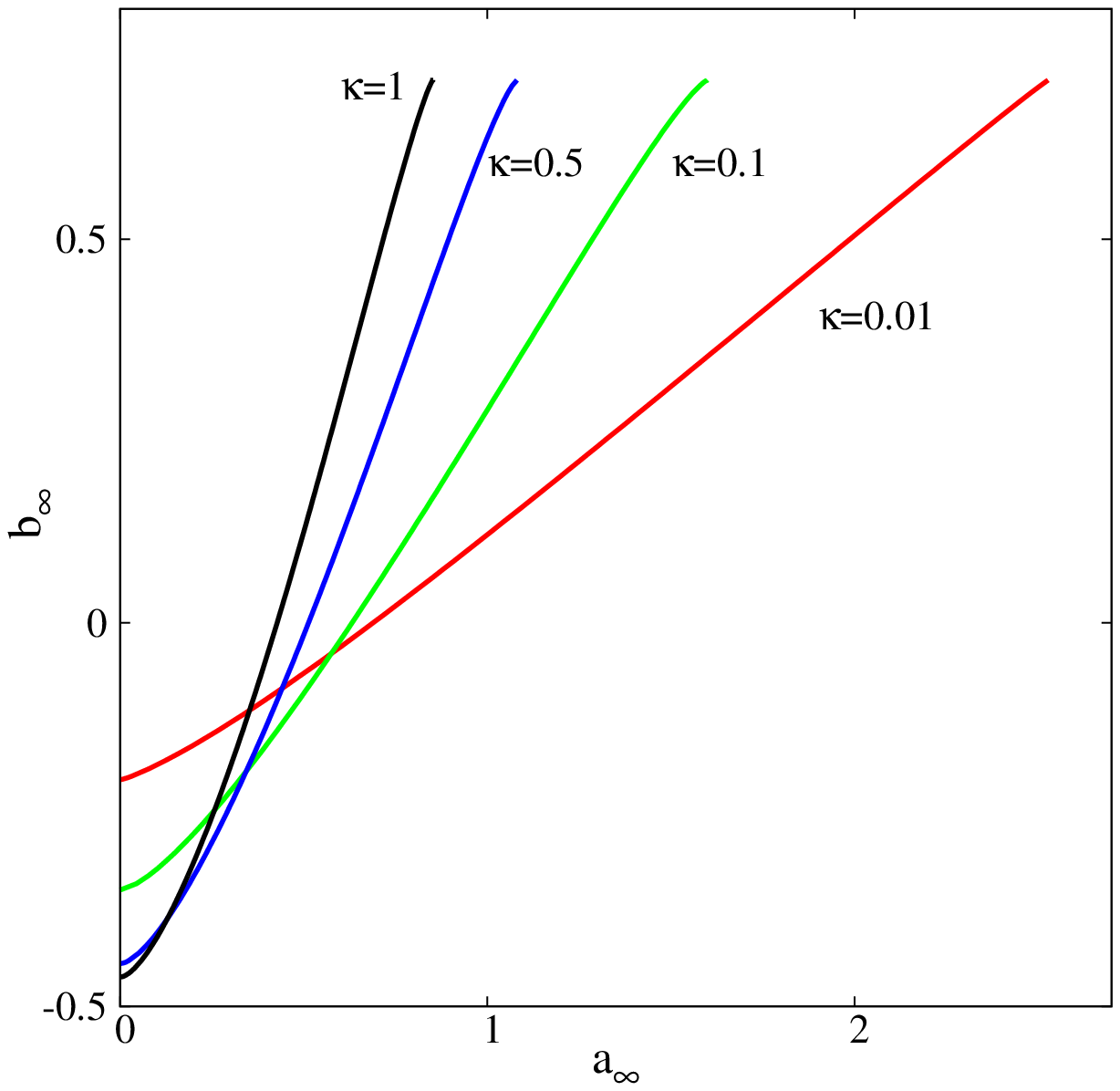}  
\includegraphics[height=2.8in]{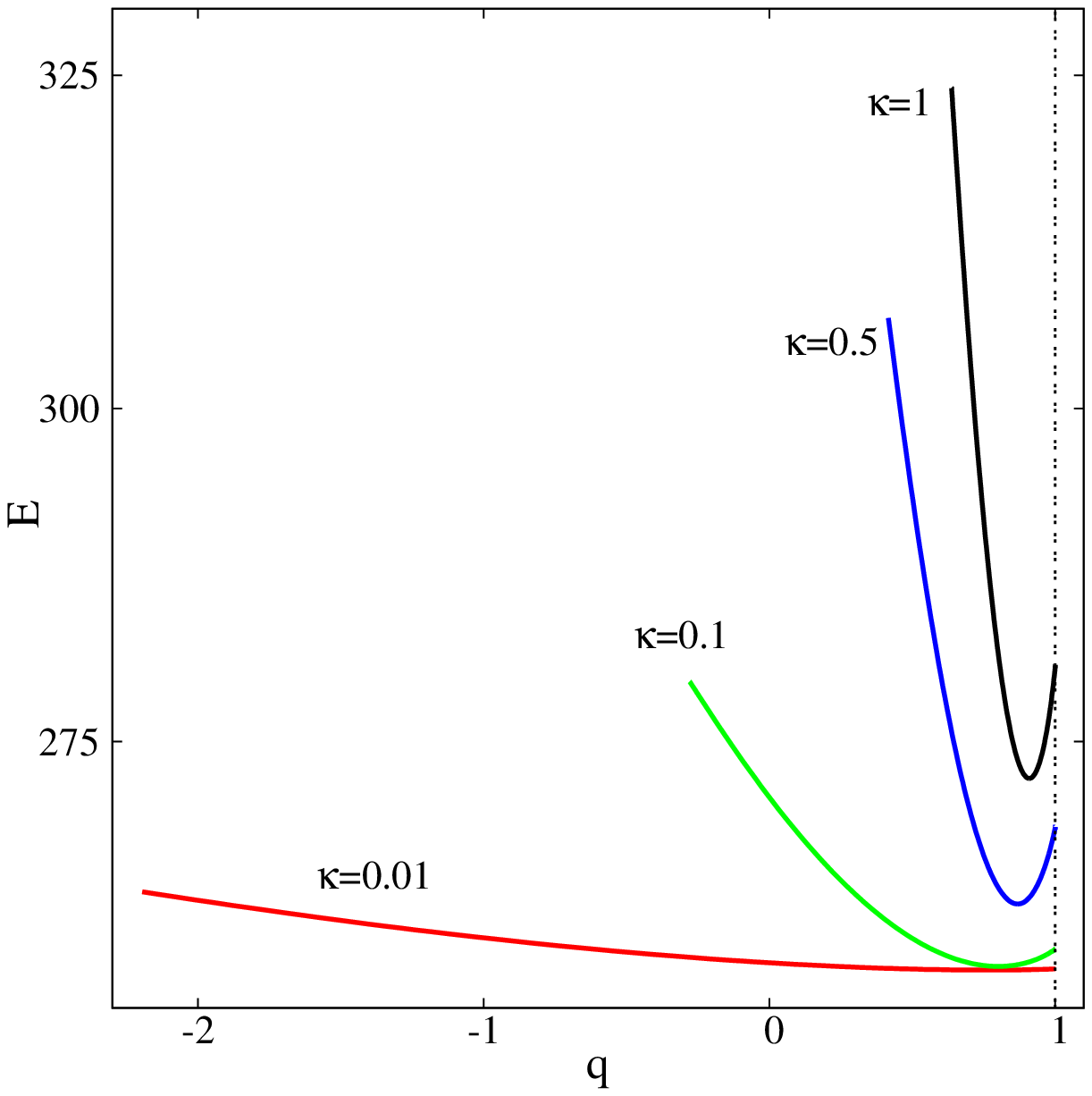}  
\caption{
{\it Left panel:}
The asymptotic value of the electric potential 
$b_\infty$
is shown 
 $vs.$
the value  
$a_\infty$ 
of the magnetic potential at infinity,
 for solutions with
 $\lambda=\lambda_1=\lambda_3=1$, $\lambda_2=0$, and $\kappa=0.01, 0.1, 0.5, 1.0$. 
{\it Right panel:}
Energy  $E$ $vs.$
 topological charge $q$ for the same solutions. 
} 
\label{fig1}
\end{figure} 

\begin{figure}[h!]
\centering
\includegraphics[height=2.8in]{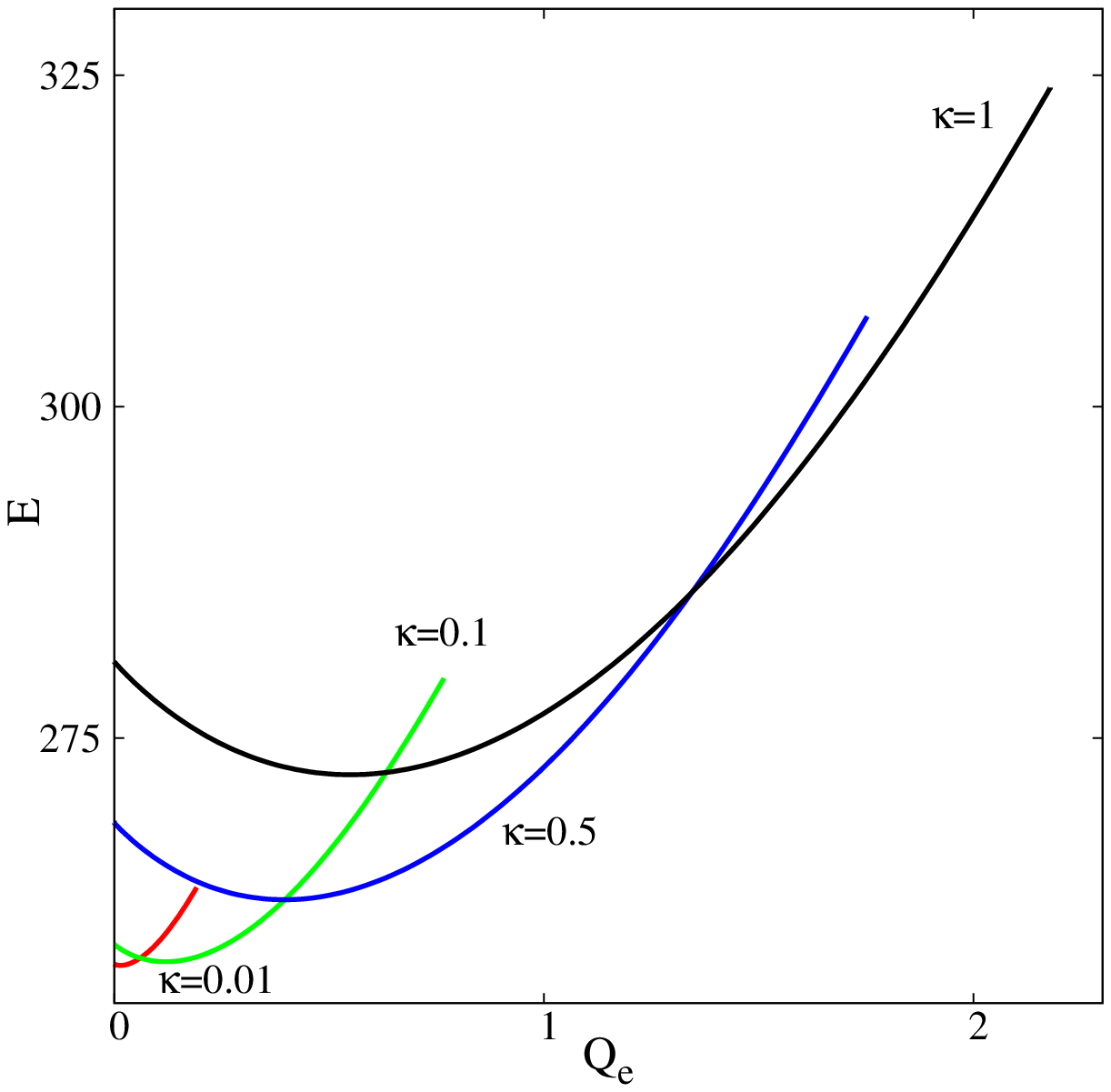}
\includegraphics[height=2.8in]{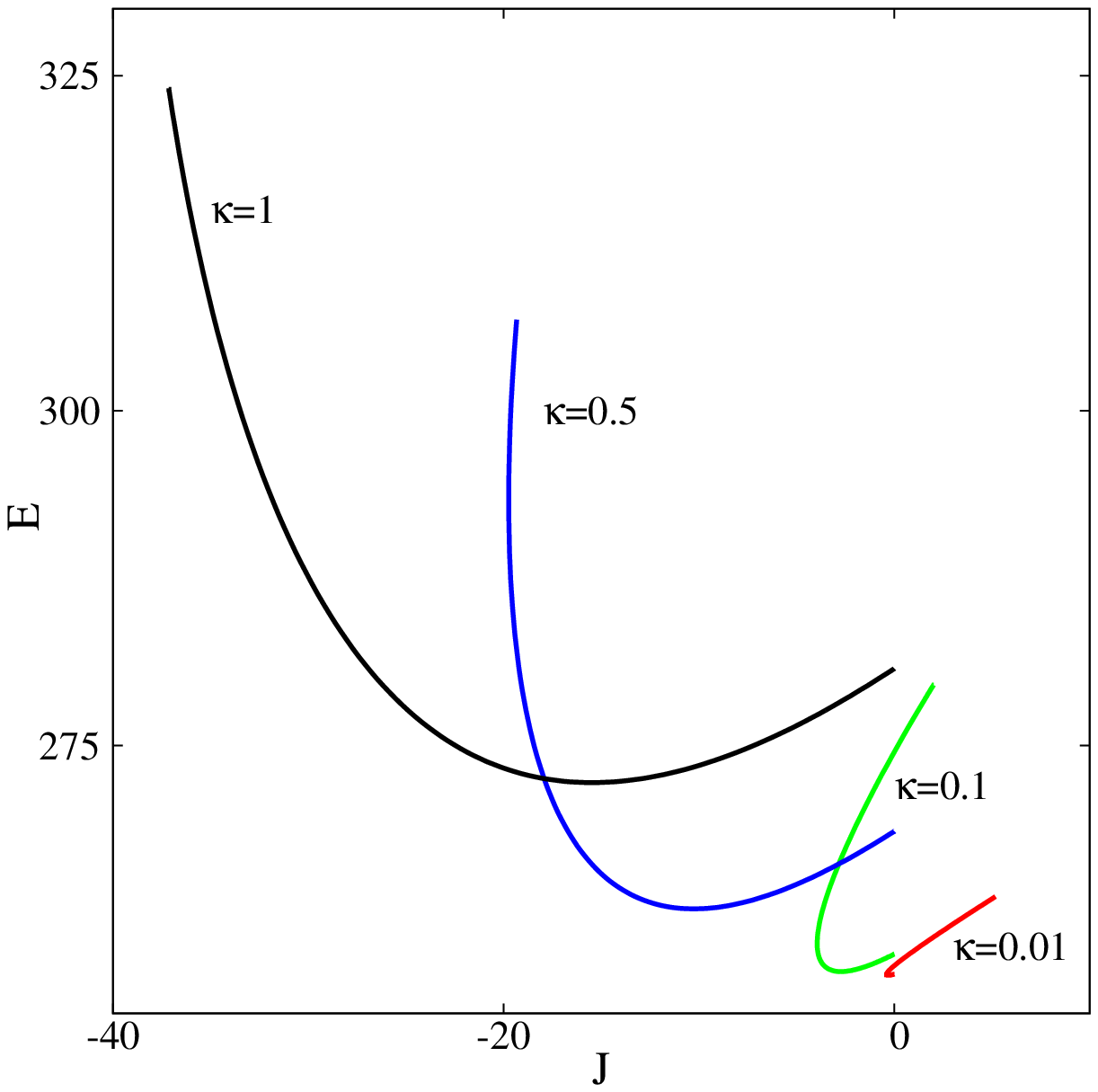}    
\caption{ 
The total mass-energy  $E$ $vs.$
is shown $vs.$
 electric charge $Q_e$ 
(left panel) and
 angular momentum $J$
(right panel) 
 for same solutions as in Fig.~\ref{fig1}.} 
\label{fig3}
\end{figure} 

 \medskip

The dependence of the mass-energy $E$ on the topological charge $q$ is exhibited in Fig.~\ref{fig1} (right panel).
While in the SO(2) gauged model in Ref. \cite{Navarro-Lerida:2018giv}
featuring a standard $F^2$ term, the   topological charge is always $q=1$,
the situation in this case is different.
While the solutions with  $a_\infty=0$ still have a unit topological charge,
a varying $a_\infty$ gives rise to solutions with $q<1$, as implied by (\ref{vr10}).
Moreover, while for large values of the CS constant one finds solutions with $q$ still positive,
decreasing $\kappa$
allows for configurations with
vanishing topological charge, negative values
 being also permitted. 
Notice that the allowed range of  $q$ decreases as $\kappa$ increases. 
Also, solutions in the limit $\kappa \to 0$ are either static and uncharged or singular.

 \medskip

The dependence of the total mass-energy $E$ on the electric charge $Q_e$ and the angular momentum $J$ is addressed 
in Figs.~\ref{fig3} (left and right panels, respectively). 
It is clearly seen that the solution with the least energy does not correspond to the uncharged solution, in general.
 That fact can be clearly seen for $\kappa=1$ curve  in the plot.
 On the other hand, the relation between the energy and the angular momentum is also quite unconventional.
 The least energetic solution corresponds usually to a rotating (with $J \neq 0$) solution. 
Moreover, for small values of $\kappa$ there exist rotating solutions
(in the sense of solutions having a non-zero angular momenta density,
$T_{\varphi_{12}}^t\neq 0$) 
with vanishing total angular momenta ($J=0$). 

We would like to point out that most of these behaviors 
also happen in $2+1$ dimensions \cite{Navarro-Lerida:2018giv}, 
so one can infer that they are not inherent to a concrete spacetime odd dimension 
but to the present of a CS term in the Lagragian, 
together with the possibility of existence of solutions with a non-vanishing $a_\infty$.

\section{Summary and outlook}

The main purpose of this work was to expose the mechanism leading 
to several new unusual features discovered in the Abelian gauged $O(3)$ Skyrme model with a Chern-Simons term in $2+1$ dimensions,
which were reported in Ref.~\cite{Navarro-Lerida:2016omj}. 
The features in question are: $(i)$ the value of the topologically invariant baryon number $n$ are altered by the gauge dynamics
as expressed by Eq. \re{q2+1}, and $(ii)$ the (global) electric charge and angular momentum of the (gauged)
Skyrmions exhibit the non-standard feature that their slopes $vs.$ the mass-energy are not only
positive, but also negative, resulting from \re{QJ2+1}.

Specifically, this mechanism hinges on the possibility of $a_\infty$, the asymptotic value of the magnetic field $a(r)$
of the gauged Skyrmion, not vanishing~\footnote{
This is exclusively a feature of gauged Skyrmions in contrast to
gauged Higgs solitons. In the latter case $a_\infty=0$ always, by virtue of the covariant constancy of the Higgs field resulting from
symmetry breaking dynamics.
}. 
A necessary ingredient of this mechanism is
the Chern-Simons dynamics. The Chern-Simons density \re{CS5stat} intertwines the magnetic field $F_{ij}$ and the
electric potential $A_0$, the latter appearing linearly. Since the asymptotic value
$b_\infty$ of $A_0$ is not fixed, then the dependence of  $a_\infty$ on $b_\infty$ 
provides the leverage on the variation of the former.

Our strategy here was to replicate the features in question for an Abelian gauged $O(5)$ Skyrme model 
featuring a $F^4$ Maxwell-like term in $d=4+1$ dimensions, endowed with the Chern-Simons density.
While this model is considerably more complex than the one~\cite{Navarro-Lerida:2016omj} 
in $d=2+1$ dimensions, the said properties are qualitatively reproduced.
This analysis exposes the studied mechanism. Moreover, it can be seen that it can be reproduced in all $d=2p+1$ dimensions
for a $O(2p+1)$ Abelian gauged Skyrme model with a $F^{2p}$ Maxwell-like term.
The choice of $F^{2p}$ Maxell-like terms is made to ensure the appropriate asymptotic decay of the Abelian field, such that
the surface integrals giving the electric charge and angular momentum
receive their contributions from the $p$-th Pontryagin density as formulated by Paul and Khare~\cite{Paul:1986ix}.
(Of course, in all but $2+1$ dimensions the Pontryagin index vanishes.)

In the $2+1$ dimensional case,  
it was shown in \cite{Navarro-Lerida:2016omj} that the dependence of the
mass-energy of the solution on the electric charge and angular momentum $Q_e$ and $J$ was a $non\ standard$ case,
in the sense that $E$ did not increase monotonically as in the $standard$ case,
with increasing $(Q_e,J)$. 
The effect of the Chern-Simons dynamics resulted in negative gradients of $E\ vs.\ (Q_e,J)$
in some regions of the parameter space. 
In \cite{Navarro-Lerida:2018siw}, it was
found that the same mechanism resulted in the departure of the ``baryon number'' $q$ of the gauged solutions
from the value of the baryon number $n$ of the Skyrmion prior to gauging.
Also,
$(Q_e,J,q)$, are evaluated as ``surface integrals''
 encoded by $a_\infty$, the asymptotic value of the magnetic potential $a(r)$ (see Eqs. (\ref{q2+1}), (\ref{QJ2+1})).

Similarly, in the 4+1 dimensional case, the  solutions are characterised by  
$a_\infty$ and $b_\infty$ -- the asymptotic value of the electric function $b(r)$. 
The Chern-Simons dynamics results in the dependence of $a_\infty$ on $b_\infty$, and the studied  family of solutions
result from the (free) choice of  $b_\infty$.
 
The results of our numerical analysis are displayed in Figures {\bf 1}, {\bf 2} and {\bf 3}:
\\

\noindent
$(i)$ The relation of $a_\infty$ with $b_\infty$, displayed in Figure {\bf 1}. 
In this case there is a detail in which the $2+1$ and $4+1$ cases differ, namely that in the former case $a_\infty$
can take both positive and negative values, while in the case at hand it takes only positive values. 
This is not surprising since in $2+1$ dimensions $a(r)$ appears in the imposition of
azimuthal symmetry in $\R^2$, while the function $a(r)$ here has a different origin, resulting from the imposition of
an enhanced symmetry on a system featuring two distinct magnetic potentials.
\\

\noindent
$(ii)$ The change (deformation) of the baryon number of the Skyrmion due to the influence of the gauge field.
This is displayed in Figure {\bf 2}, by plotting the deformed ``baryon number'' $vs.$
the energy $E$ of the continuum of solutions.  
\\

\noindent
$(iii)$ The $unusual$ dependence of $E$ on $Q_e$ and $J$ involving both negative and positive gradients,
displayed in {\it left panel} of Figure {\bf 3} and in {\it right panel} of Figure {\bf 3},
respectively.
 
\medskip

It may be instructive to display the Lagrangians in $2p+1$ dimensions, generalising the $p=2$ Lagrangian \re{L1}. 
This can be expressed formally as
\be
\label{Lp}
{\cal L}= \frac{1}{(2p)!}\la_M F(2p)^2+\ka\,\Om_{\rm CS}^{(2p+1)}-\la_1|\f(1)|^2+\la_2|\f(2)|^2-\dots+\la_{2p}|\f(2p)|^2+\la V[\f^{2p+1}]
\ee
where $F(2p)$ is the $p$-fold totally antisymmetrised product of the $2$-form Maxwell curvature,
$\Om_{\rm CS}^{(2p+1)}$ is the Abelian Chern-Simons density
\bea
\Om_{\rm CS}^{(2p+1)}&=&-\frac12\vep^{\lambda \mu_1\mu_2\dots\mu_{2p-1}\mu_{2p}}A_{\la}\,F_{\mu_1\mu_2}\dots F_{\mu_{2p-1}\mu_{2p}}\label{CSp}\\
\Om_{\rm static}^{(2p+1))}&=&-\vep^{i_1i_2\dots i_{2p-1}i_{2p}}A_{0}\,F_{i_1i_2}\dots F_{\mu_{2p-1}\mu_{2p}}\ ,
\quad i_1,i_2,..=1,2,\dots,2p-1,2p\,,\label{CSpstat}
\eea
where, again, in the static limit \re{CSpstat},
 the linear dependence on $A_0$ implies the dependence of $a_\infty$ on $b_\infty$.
The $|\f(p)|^2$ are the Skyrme kinetic terms consistent with Derrick scaling, expressed in terms of $\f(p)$ which is
the $p$-fold totally antisymmetrised product of the $2$-form $D_\mu\f^a$. $V[\f^{2p+1}]$ is the (optional) potential term.

To keep the full analogy with the considered $p=1,2$ models, 
we impose first a $p$-azimuthal symmetry ($i.e.$ the solutions 
possess an azimuthal symmetry in
each $\R^2$ sub-plane of $\R^{2p}$), 
and then an enhanced symmetry, by
equating the angular momenta in each of the $p$ sub-planes, 
$J_1=J_2=\dots=J_{p}\equiv J$
 (see  Ref. \cite{Navarro-Lerida:2020jft} for a discussion 
of this procedure in $d=5$ dimensions). 
In other words, a suitable Maxwell-Skyrme Ansatz factorizing the angular dependence exists in the 
generic $d=2p+1$ case, which generalizes the expressions (\ref{n11}), (\ref{n12}).
As such, the $p-$model reduces to an effectively one dimensional radial system,
in terms of three functions $f(r),a(r)$ and $b(r) $.

The following speculative comparisons of the quantities $Q_e(p), J(p)$ and $q(p)$, in dimensions $d=2p+1$ may be illustrative:
\
\bea
&&Q_e(1)\sim n-a_\infty\ ,\ \ \ Q_e(2)\sim \,a_\infty^2\ \qquad\quad \Rightarrow\ \  Q_e(p)\sim \,a_\infty^p\nonumber\\
&&J(1)\sim\pi(a_\infty^2-n^2)\ ,\ \ J(2)\sim\pi^2(3a_\infty^3-2a_\infty)\ \Rightarrow\ J(p)\sim\pi^p(\al_1a_\infty^{p+1}+\al_2a_\infty^{p-1}+\dots)\nonumber\\
&&q(1)\sim n+a_\infty \ ,\ \ q(2)\sim  1-\frac12\,a_\infty^2 \ \quad \Rightarrow\ \ q(p)\sim 1-\bt\,a_\infty^p \label{spec}
\eea
where the constants $\al_1,\al_2,\dots$ and $\bt$ have only symbolic meaning.

\medskip
What distinguishes the $p=1$ members in \re{spec} is that they feature the {\it winding number} $n$,
 which is absent in the $p>1$ case. The winding number $n$ is in fact the $1$st
Ponytryagin number in the expression of the electric charge for the $p=1$ counterpart of \re{electch} above,
\be
Q_e(1)=\frac{1}{4\pi}\int j_0\,d^2x=\frac{1}{8\pi}\int\left(\pa_i F_{i0}\label{Q1}
-\frac12\ka\vep^{ij}\,F_{ij}\right)d^2x\,,
\ee
$n$ resulting from the integral of the second term in the integrand of \re{Q1}.
For $a_\infty= 0$, $Q_e(1)$ is ``quantised'' with $n$, as in Refs.~\cite{Paul:1986ix,Hong:1990yh}.
Clearly, in $4+1$ dimensions the $2$nd Pontryagin integral in \re{electch} vanishes for the Abelian curvature, unless if $a_\infty\neq 0$.

The electric charge for the general system \re{Lp} is
\be
\label{Qp}
Q_e(p)=\frac{1}{2^{p+1}\pi}\int\left(F_{j_1j_2\dots j_{2p-2}}\pa_i F_{i0j_1j_2\dots j_{2p-2}}
-\frac12\ka\vep^{j_1j_2\dots j_{2p}}\,F_{j_1j_2\dots j_{2p}}\right)d^{2p}x\,.
\ee


The first term in the integrand will decay sufficiently fast that it will not contribute to the integral.
The contribution of the second term will also vanish unless $a_\infty\neq 0$.

For $Q_e(p)$ to receive a contribution from the $p$-th Pontryagin integral ($p\ge 2$) the Chern-Simons density must feature a non-Abelian field.
The natural choice for such a system in
$2p+1$ dimensions is a $SO_{\pm}(2p)\times SO(2)$ gauge theory.

In the $p=2$ case at hand, such a candidate is a $SO_{\pm}(4)\times SO(2)$
(or $SU(2)\times U(1)$) model which may support solutions with nonvanishing $2$nd Pontryagin
integral. For such putative solutions $Q_e(2)$ cannot be ``quantised'' by the $2$nd Pontryagin charge as
is the case in $2+1$ dimensions~\cite{Paul:1986ix} where $Q_e(1)$ is quantised by the $1$st Pontryagin charge.
This is because in the $p=2$ case the $SU(2)$ Yang-Mills equation is sourced by the Abelian field, $i.e.$, it is not the vacuum
Yang-Mills equation solved by the self-dual (BPST) field. Such a $4+1$ dimensional model is under study.

\section*{Acknowledgements}
We would like to thank Valery Rubakov for important and very useful discussions.
The  work of E.R.  is  supported  by  the Center  for  Research  and  Development  in  Mathematics  and  Applications  (CIDMA)  
through  the Portuguese Foundation for Science and Technology (FCT - Fundacao para a Ci\^encia e a Tecnologia), 
references UIDB/04106/2020 and UIDP/04106/2020 and by national funds (OE), through FCT, I.P., 
in the scope of the framework contract foreseen in the numbers 4, 5 and 6 of the article 23,of the Decree-Law 57/2016, of August 29, changed by Law 57/2017, of July 19.  
We acknowledge support  from  the  projects  PTDC/FIS-OUT/28407/2017 
 and  CERN/FIS-PAR/0027/2019.  
 This work has further been supported by the European Union’s Horizon 2020 research and innovation (RISE) programme H2020-MSCA-RISE-2017 Grant No. FunFiCO-777740.  
The authors would like to acknowledge networking support by the COST Action CA16104

\begin{small}

\end{small}


\begin{thebibliography}{99}
\bibitem{Skyrme:1961vq}
  T.~H.~R.~Skyrme,
  Proc.\ Roy.\ Soc.\ Lond.\ A {\bf 260} (1961) 127.
\bibitem{Skyrme:1962vh}
  T.~H.~R.~Skyrme,
  Nucl.\ Phys.\  {\bf 31} (1962) 556.
\bibitem{Tchrakian:2015pka}
  D.~H.~Tchrakian,
  J.\ Phys.\ A {\bf 48} (2015) no.37,  375401
  [arXiv:1505.05344 [hep-th]].
\bibitem{Piette:1994jt}
B.~M.~A.~G.~Piette, W.~J.~Zakrzewski, H.~J.~W.~Mueller-Kirsten and D.~H.~Tchrakian,
Phys. Lett. B \textbf{320} (1994), 294-298.
\bibitem{Brihaye:2017wqa}
Y.~Brihaye, C.~Herdeiro, E.~Radu and D.~H.~Tchrakian,
JHEP \textbf{11} (2017), 037
[arXiv:1710.03833 [gr-qc]].
\bibitem{Tchrakian:1997sj}
D.~H. Tchrakian,
Lett. Math. Phys. \textbf{40} (1997), 191-201.
\bibitem{Tchrakian:2002ti}
D. H.~Tchrakian,
 {\it ``Winding number versus Chern-Pontryagin charge,''}
[arXiv:hep-th/0204040 [hep-th]].
\bibitem{Callan:1983nx}
C.~G.~Callan, Jr. and E.~Witten,
Nucl. Phys. B \textbf{239} (1984), 161-176.
\bibitem{Tchrakian:2010ar}
  T.~Tchrakian,
  J.\ Phys.\ A {\bf 44} (2011) 343001
  [arXiv:1009.3790 [hep-th]].


\bibitem{Navarro-Lerida:2016omj}
  F.~Navarro-L\'erida, E.~Radu and D.~H.~Tchrakian,
  Phys.\ Rev.\ D {\bf 95} (2017) no.8,  085016
  [arXiv:1612.05835 [hep-th]].
\bibitem{Navarro-Lerida:2018giv}
F.~Navarro-L\'erida, E.~Radu and D.~Tchrakian,
Phys. Lett. B \textbf{791} (2019), 287-292
[arXiv:1811.09535 [hep-th]].
\bibitem{Navarro-Lerida:2018siw}
  F.~Navarro-L\'erida and D.~H.~Tchrakian,
  Phys.\ Rev.\ D {\bf 99} (2019) no.4,  045007
  [arXiv:1812.03147 [hep-th]].


\bibitem{Navarro-Lerida:2020jft}
F.~Navarro-L\'erida, E.~Radu and D.~Tchrakian,
Phys.\ Rev.\ D {\bf 101} (2020) no.12, 125014
[arXiv:2003.05899 [hep-th]].


\bibitem{Paul:1986ix}
  S.~K.~Paul and A.~Khare,
  Phys.\ Lett.\ B {\bf 174} (1986) 420
   [Erratum-ibid.\  {\bf 177B} (1986) 453].
\bibitem{Hong:1990yh}
  J.~Hong, Y.~Kim and P.~Y.~Pac,
  Phys.\ Rev.\ Lett.\  {\bf 64} (1990) 2230;
	\\
  R.~Jackiw and E.~J.~Weinberg,
  Phys.\ Rev.\ Lett.\  {\bf 64} (1990) 2234.
\bibitem{Ghosh:1995ze}
  P.~K.~Ghosh and S.~K.~Ghosh,
  Phys.\ Lett.\ B {\bf 366} (1996) 199
  [hep-th/9507015];
	\\
  K.~Kimm, K.~-M.~Lee and T.~Lee,
  Phys.\ Rev.\ D {\bf 53} (1996) 4436
  [hep-th/9510141];
	\\
  K.~Arthur, D.~H.~Tchrakian and Y.~-s.~Yang,
  Phys.\ Rev.\ D {\bf 54} (1996) 5245.
 
\bibitem{VanderBij:2001nm}
J.~Van der Bij and E.~Radu,
Int. J. Mod. Phys. A \textbf{17} (2002), 1477-1490
[arXiv:gr-qc/0111046 [gr-qc]].
\bibitem{Tseytlin:1997csa}
  A.~A.~Tseytlin,
  Nucl.\ Phys.\  B {\bf 501} (1997) 41
  [arXiv:hep-th/9701125].
\bibitem{Polchinski:1998rr}
  J.~Polchinski,
  {\it `String theory. Vol. 2: Superstring theory and beyond'}, 
Cambridge, Cambridge University Press, (1998).
\bibitem{Blazquez-Salcedo:2017cqm}
J.~L.~Bl\'azquez-Salcedo, J.~Kunz, F.~Navarro-L\'erida and E.~Radu,
Phys. Lett. B \textbf{771} (2017), 52-58
[arXiv:1703.04163 [gr-qc]].
\bibitem{Blazquez-Salcedo:2017kig}
J.~L.~Bl\'azquez-Salcedo, J.~Kunz, F.~Navarro-L\'erida and E.~Radu,
Phys. Rev. D \textbf{97} (2018) no.8, 081502
[arXiv:1711.08292 [gr-qc]].
\\
J.~L.~Bl\'azquez-Salcedo, J.~Kunz, F.~Navarro-L\'erida and E.~Radu,
JHEP \textbf{02} (2018), 061
[arXiv:1711.10483 [gr-qc]].


\end{thebibliography}
\end{document}